\documentclass[12pt]{article}
\usepackage{a4wide}
\usepackage{cite}
\usepackage{pifont}
\usepackage{amsmath}
\ifx\pdfoutput\undefined
\usepackage{hyperref}
\usepackage{graphicx}
\else
\usepackage[pdftex]{hyperref}
\usepackage[pdftex]{graphicx}
\fi
\graphicspath{{./}{./figures/}}

\newcommand{\FMslash}[1]{\not #1}

\def\bbuildrel#1_#2^#3%
{\mathrel{\mathop{\kern 0pt#1}\limits_{#2}^{#3}}}

\newcommand{\ovl}[1]{\overline{#1}}
\newcommand{\ice}[1]{\relax}
\newcommand{\re}[1]{(\ref{#1})}

\newcommand{\beq}{\begin{equation}}
\newcommand{\eeq}{\end{equation}}
\newcommand{\bea}{\begin{eqnarray}}
\newcommand{\eea}{\end{eqnarray}}

\newcommand{\ba}{\begin{array}} 
\newcommand{\ea}{\end{array}}

\newcommand{\as}{a_s}

\newcommand{\als}{\alpha_s}

\newcommand{\msbar}{{\scriptsize \overline{\rm MS}}}

\let\*=\,

\begin{document}    

\begin{titlepage}
\noindent
\hfill SFB/CPP--09--104\\
\mbox{}
\hfill TTP09--40\\
\mbox{}

\vspace{0.5cm}
\begin{center}
  \begin{Large}
    \begin{bf}
Wilson Expansion of QCD Propagators at Three  Loops:
Operators of Dimension Two and Three
    \end{bf}
  \end{Large}
  \vspace{0.8cm}

  \begin{large}
    K.G. Chetyrkin\footnote{Permanent address: Institute
    for Nuclear Research, Russian Academy of Sciences, Moscow 117312, Russia.} {\normalsize and }
    A. Maier
  \end{large}
  \vskip .7cm
	 {\small {\em 
	    Institut f{\"u}r Theoretische Teilchenphysik,
            Universit{\"a}t Karlsruhe,
            D-76128 Karlsruhe, Germany}}
	 \vskip .3cm
{\bf Abstract}
\end{center}
\begin{quotation}
  \noindent
  In this paper we construct the Wilson short distance operator product expansion
  for the gluon, quark and ghost  propagators in QCD, including   operators of dimension two  and three, namely,
 $A^2$, $m^2$, $m \,A^2$, $\ovl{\psi} \,\psi$   and $m^3$.
  We compute analytically the coefficient functions of these  operators 
  at  three loops for all three propagators in the general covariant gauge. 
  Our results, taken in the Landau gauge, should help to improve the accuracy of extracting the 
  vacuum expectation values of these operators from lattice simulation of the QCD propagators.  
\end{quotation}

\end{titlepage}

\ice{
\begin{verbatim}
                               Plan 

1. Intro

2. OPE for two-point field correlators

2.1 Generalities

2.2  Ghost propagator

2.3  Gluon propagator

2.4  Quark propagator 

\end{verbatim}
}

\section{Introduction\label{sec:Introduction}}

Two-point correlation functions of the fundamental fields of the QCD
Lagrangian -- that is gluon, ghost and quark propagators -- are of
direct importance in any perturbative treatment of QCD. Suffice it to
say that the corresponding wave function renormalization constants are
vital ingredients in calculations of the QCD $\beta$-function and the
quark mass anomalous dimensions (currently known at four-loop level
\cite{vanRitbergen:1997va,Czakon:2004bu,Vermaseren:1997fq,Chetyrkin:1997dh}).
Scheme-invariant versions of these propagators are presently known in
NNNLO (that is up to and including three loops) in arbitrary covariant
gauge \cite{Chetyrkin:2004mf} including the Landau one, which is distinguished from the
point of view of lattice simulations.

Purely perturbative treatment essentially assumes a weak-coupling regime. 
QCD propagators, especially the gluon and the quark ones, have been
much under examination also beyond perturbation theory (that is in the
strong-coupling regime).  Here one should mention at least two broad
directions, namely, the use of Schwinger-Dyson equations 
(for reviews see e.g., \cite{Roberts:1994dr,Maris:2003vk,Alkofer:2000wg}) 
 and non-perturbative computation on the lattice by Monte Carlo
simulations.  In what follows we will concentrate our discussion on
the latter.

It is expected --- due to the asymptotic freedom --- that the behavior
of full QCD-propagators is to be governed at sufficiently large
momentum transfers by perturbation theory {\em and} by the Operator
Product Expansion (OPE)
\cite{Wilson:1969zs,Politzer:1976tv,Shifman:1978bx}.  Thus, by
comparing results of continuum perturbation theory calculations
with those of lattice simulations one hopes to get a lot of
information about the (renormalized) running coupling constant and
quark masses as well as on {\em condensates} --- Vacuum Expectation
Values (VEV's) of composite operators --- entering into OPE.  The
idea has been pursued  in lattice simulations performed by various groups.
(As for investigating condensates in lattice framework along these
lines, see, e.g.
Refs.~\cite{RuizArriola:2004en,Gimenez:2005nt,Cucchieri:2006xi,Boucaud:2008gn}
and  also references therein for earlier results and for more
lattice-specific information).


While purely perturbative contributions to the QCD propagators have
been computed in NNNLO, the corresponding (power suppressed)
condensate contributions are usually known only at leading order or,
at best, at next-to-leading order. To be specific, let us consider the
gluon and ghost propagators in Landau gauge (for space-like momentum $q^2 < 0$)
\beq
D^{ab}_{\mu\nu}(q) = \frac{\delta^{ab}}{-q^2}
\left[
 -g_{\mu\nu} + \frac{q_{\mu} q_{\nu}}{q^2}
\right] D^g(Q)
{},
\ \
\Delta^{ab}(q) = \frac{\delta^{ab}}{-q^2}
\ \
 D^h(Q)
\label{gluon_ghost_prop}
{}
\eeq
with $Q = \sqrt{ -q^2}$.
The 
{\em dressing functions} $D^g$ and  $D^h$ can be decomposed in terms of 
the appropriate OPE as follows 
\beq
D^?(Q^2) \bbuildrel{=\!=\!=}_{Q^2 \to \infty}^{} 
D^?_0(\mu/Q,\as) + \sum_i \frac{C^?_i(\mu/Q,\as)}{Q^{\,d_i}}
  \, \langle O_i \rangle
{},
\label{gl+gh:OPE}
\eeq
where $?$ stand for $g$ or $h$, $ D^?_0(\mu/Q,\as)$ is the purely perturbative contribution, $\mu$
is the renormalization scale and $\as = \frac{\als}{\pi}=
\frac{g_s^2}{4\,\pi^2}$ is the quark-gluon coupling constant.  The sum goes over all scalar operators with
vacuum quantum numbers; $d_i$ stands for the dimension of the
operator $O_i$ in  mass units.

Assuming the case of massless QCD, the leading non-perturbative corrections in \re{gl+gh:OPE}
should come from operators with the lowest possible mass dimension $d_i =2$, 
namely (to be in agreement with the commonly used in lattice publication sign convention
we effectively  use below the {\em euclidean} scalar product in 
the definition of $A^2$)
\[
A^2 \equiv -A^a_{\mu} A^{a\,{\mu}} \ \ \ \mbox{and} \ \ \ i\,\ovl{C}^a  C^a
{},
\label{dim:2}
\]
where $A^{a}_{\mu}$ is the gauge  field , $C(\ovl{C})$ is  the ghost (antighost) field.
Within the class of covariant gauges\footnote{
By a covariant gauge 
we mean the one generated by adding the term $-\frac{1}{2\xi_L}(\partial_{\nu} A^a_{\mu}) (d_{\nu} A^a_{\mu})$
to the  gauge invariant Yang-Mills Lagrangian; the corresponding expression for
the tree-level   vector boson propagator reads 
$\frac{\delta^{ab}}{-q^2}
\left[
 -g_{\mu\nu} + (1- \xi_L)\frac{q_{\mu} q_{\nu}}{q^2}
\right]
;
$ the choice of the  Landau gauge corresponds to \mbox{limit of $\xi_L \to 0$}.}
the coefficient function of the second operator is  known to vanish identically  in every   OPE 
\cite{Lavelle:1988eg}. In what follows we will not consider this operator. 

\ice{On the other hand,}The first operator, the {\em gluon mass condensate}\footnote{ We  use this
expression as  the title of just {\em  gluon  condensate} is traditionally referred to the VEV of the 
operator 
$ G^a_{\mu\nu}G^a_{\mu\nu} $ starting from the seminal works by the ITEP group \cite{Shifman:1978bx}.}, 
does have  nonzero
coefficient functions  $C^?_{A^2}$ already in the tree approximation, namely
\cite{Lavelle:1988eg} (see, also \cite{Boucaud:2000nd,Boucaud:2001st,Boucaud:2005xn})
\beq
C^g_{A^2} = g_s^2\,\frac{3}{32} + {\cal{O}}(\as^2)
{},
\\
\ \ \ 
C^h_{A^2} = g_s^2\,\frac{3}{32} + {\cal{O}}(\as^2)
\label{CF:g:gh:tree}
{}.
\eeq
From the phenomenological side, lattice simulations carried on by
Boucaud et al in a series of publications
\cite{Boucaud:2000nd,Boucaud:2001st,Boucaud:2002nc,Boucaud:2003xi,Boucaud:2005rm,Boucaud:2008gn}
(see, also \cite{DeSoto:2001qx}) seem to demonstrate the existence of
effects of order $1/Q^2$ in both gluon and ghost dressing
functions\footnote{The gluon mass condensate as well as the quark
condensate also show up  in the quark propagator
\cite{Lavelle:1992yh,RuizArriola:2004en,Gimenez:2005nt,Bowman:2004xi,Boucaud:2005rm,Bowman:2006zk}, 
see Section 3  below.}.  Moreover,
numerical fits produce the results consistent with the OPE description
of power suppressed $1/Q^2$ corrections to the gluon, ghost and quark
propagators.  This means, for instance, that {\em one and the same
value} of the gluon mass condensate
\cite{Boucaud:2008gn}
\beq
g^2_s\, \langle A^2 \rangle  = 5.1 ^{+0.7}_{-1.1} \,\,\mbox{GeV}^2
\eeq
multiplied by the {\em tree level} CF's \re{CF:g:gh:tree} together
with purely perturbative contributions  (known to {\em three loops}) 
describe 
the ghost and gluon dressing functions  over 
the whole available momentum window $ 2\,\,  \mbox{GeV} \, \le Q \, \le 6\,\, \mbox{GeV}$.

On the other hand,  a study of a dressing function itself
could, obviously, at best result in the determination of the product of
the CF and the VEV of a composite operator (even if one assumes no
contamination from operators of higher mass dimension). Thus,
knowledge of higher order corrections to the coefficient
functions of condensates is of some importance, at least for better
understanding the results of lattice simulations.

The quarks are massive. As a consequence the possible
composite operators could contain powers of quark masses along with
quantum fields. It is worthwhile to remember at this point that in
``good'' renormalization schemes like those based on the dimensional
regularization \cite{Cicuta:1972jf,Ashmore:1972uj,'tHooft:1972fi} and
minimal subtractions \cite{'tHooft:1973mm} the coefficient functions
of any (short distance) OPE obey the following important
property\footnote{To our knowledge it was first established in
\cite{Chetyrkin:1982zq}; see also \cite{Tkachov:1983st}.}: their dependence on any particle/field
masses is {\em polynomial}. In particular, it means that any more
complicated mass dependence of a correlator will be ``hidden'' in the
corresponding VEV of composite operators. It also means that if one
allows, as we do, mass factors to be used in constructing composite
operators, then their coefficient functions become totally
mass-independent by definition. Limiting ourselves to operators with
mass dimensions not higher than three we arrive to the following list
of operators which could appear in OPE for the QCD propagators:
\begin{equation}
A^2 \equiv  A^a_{\mu} A^a_{\mu},\ \  m^2, \ \ \ m^3, \ \ m\,A^2,\ \ \bar{\psi}\psi 
\label{operators}
{},
\end{equation}
where $m$ is a  quark mass.
 and we have assumed QCD with $n_f=n_l+1$ total number of quark flavors, one of those
having a mass $m$, while all others are strictly massless\footnote{
Later, in Appendix A,  we generalize our results for the case of arbitrary many massive quarks.}.

The aim of the research we are going to present was to compute the
higher order contributions (up to and including three loops ) to
coefficient functions of operators \re{operators} appearing in the OPE of QCD
propagators. The structure of the paper is as follows.  In the next
two sections we describe our results for OPE of QCD propagators.  In
the fourth section we consider the RG evolution equations for
propagators and operators under consideration and construct the scale
and scheme invariant combinations of operators and coefficient
functions. Due to their scheme independence the latter should be most
convenient for comparisons with the results of lattice calculations.
Then we briefly discuss \mbox{(in Section 5)} some technical
details of the calculations as well as software/hardware tools
employed.  Finally, a short summary of our findings is given in the
Conclusion \mbox{(Section 6)}.

We finish the  introduction by adding that in recent years, starting from works 
\cite{Gubarev:2000eu,Gubarev:2000nz}, 
the condensates of mass dimension two, especially the gluon mass condensate, have been
intensively studied in view of better understanding of confinement in
Yang-Mills theories and QCD. (For example, see recent works 
\cite{Kondo:2005zs,Kondo:2006ih,Baranov:2006vh,Arriola:2006sv,Dudal:2006tp, Sorella:2006pj,Gracey:2006dr,Capri:2006ne,Chernodub:2008kf}
and references therein). Unfortunately, any discussion of these
developments is beyond the scope of the present paper.

\section{OPE for the gluon and ghost propagators}
On dimensional grounds, the operators of dimensions three do not contribute to 
the OPE for the gluon and ghost propagators. The remaining  coefficient functions
$C^?_{m^2}$ and $C^?_{A^2}$ read:
\begin{align}
\label{eq:c_g_m2}
 C_{m^2}^g=&a_s\bigg[1+a_s\bigg(\frac{383}{24}+\frac{3}{2}\zeta_3-\frac{5}{9}n_f+\frac{93}{16}l_{\mu Q}-\frac{1}{3}l_{\mu Q}n_f\bigg)\notag\\
 &+a_s^2\bigg(\frac{7370507}{27648}-\frac{27}{64}\zeta_4-\frac{22615}{864}\zeta_5+\frac{415679}{6912}\zeta_3-\frac{69941}{3456}n_f\notag\\
 &-\frac{113}{24}n_f\zeta_3+\frac{25}{108}n_f^2+\frac{7405}{48}l_{\mu Q}+\frac{411}{32}l_{\mu Q}\zeta_3-\frac{4123}{288}l_{\mu Q}n_f\notag\\
 &-\frac{3}{4}l_{\mu Q}n_f\zeta_3+\frac{5}{18}l_{\mu Q}n_f^2+\frac{13263}{512}l_{\mu Q}^2-\frac{281}{96}l_{\mu Q}^2n_f+\frac{1}{12}l_{\mu Q}^2n_f^2\bigg)\bigg]\,,\displaybreak[0]\\
\label{eq:c_g_A2}
 C_{A^2}^g=&\frac{3}{8}\pi^2 a_s\bigg[1+a_s\bigg(\frac{785}{96}-\frac{11}{18}n_f+\frac{35}{16}l_{\mu Q}-\frac{1}{6}l_{\mu Q}n_f\bigg)\notag\\
 &+a_s^2\bigg(\frac{799087}{9216}+\frac{27}{128}\zeta_3-\frac{90371}{6912}n_f-\frac{11}{24}n_f\zeta_3+\frac{121}{324}n_f^2+\frac{70097}{1536}l_{\mu Q}\notag\\
 &-\frac{3719}{576}l_{\mu Q}n_f+\frac{11}{54}l_{\mu Q}n_f^2+\frac{2765}{512}l_{\mu Q}^2-\frac{149}{192}l_{\mu Q}^2n_f+\frac{1}{36}l_{\mu Q}^2n_f^2\bigg)\notag\\
 &+a_s^3\bigg(\frac{985590473}{884736}-\frac{243}{4096}\zeta_4-\frac{4545}{128}\zeta_5-\frac{57399}{8192}\zeta_3-\frac{159678799}{663552}n_f\notag\\
 &+\frac{33}{256}n_f\zeta_4+\frac{3355}{576}n_f\zeta_5-\frac{36455}{6912}n_f\zeta_3+\frac{1702769}{124416}n_f^2+\frac{29}{72}n_f^2\zeta_3\notag\\
 &-\frac{1115}{5832}n_f^3+\frac{38346881}{49152}l_{\mu Q}+\frac{1539}{1024}l_{\mu Q}\zeta_3-\frac{6165035}{36864}l_{\mu Q}n_f\notag\\
 &-\frac{863}{256}l_{\mu Q}n_f\zeta_3+\frac{48095}{4608}l_{\mu Q}n_f^2+\frac{11}{48}l_{\mu Q}n_f^2\zeta_3-\frac{121}{648}l_{\mu Q}n_f^3\notag\\
 &+\frac{3082507}{16384}l_{\mu Q}^2-\frac{238649}{6144}l_{\mu Q}^2n_f+\frac{1453}{576}l_{\mu Q}^2n_f^2-\frac{11}{216}l_{\mu Q}^2n_f^3\notag\\
 &+\frac{113365}{8192}l_{\mu Q}^3-\frac{1479}{512}l_{\mu Q}^3n_f+\frac{77}{384}l_{\mu Q}^3n_f^2-\frac{1}{216}l_{\mu Q}^3n_f^3\bigg)\bigg]\,,\displaybreak[0]\\
\label{eq:c_h_m2}
 C_{m^2}^h=&-\frac{3}{8}a_s^2\bigg[1+\frac{3}{2}l_{\mu Q}+a_s\bigg(\frac{35501}{1152}+\frac{9}{8}\zeta_4+\frac{15}{4}\zeta_5-\frac{35}{16}\zeta_3-\frac{23}{24}n_f\notag\\
 &+\frac{1847}{64}l_{\mu Q}+\frac{9}{4}l_{\mu Q}\zeta_3-\frac{7}{12}l_{\mu Q}n_f+\frac{441}{64}l_{\mu Q}^2-\frac{1}{4}l_{\mu Q}^2n_f\bigg)\bigg]\,,\displaybreak[0]\\
\label{eq:c_h_A2}
 C_{A^2}^h=&\frac{3}{8}\pi^2a_s\bigg[1+a_s\bigg(\frac{15}{4}+\frac{9}{8}l_{\mu Q}\bigg)\notag\\
 &+a_s^2\bigg(\frac{14853}{512}+\frac{27}{32}\zeta_3-\frac{187}{128}n_f+\frac{2145}{128}l_{\mu Q}-\frac{25}{32}l_{\mu Q}n_f+\frac{279}{128}l_{\mu Q}^2-\frac{3}{32}l_{\mu Q}^2n_f\bigg)\notag\\
 &+a_s^3\bigg(\frac{12444649}{36864}+\frac{243}{2048}\zeta_4-\frac{56745}{4096}\zeta_5+\frac{53823}{4096}\zeta_3-\frac{505459}{13824}n_f\notag\\
 &-\frac{33}{128}n_f\zeta_4-\frac{307}{256}n_f\zeta_3+\frac{13081}{20736}n_f^2+\frac{1}{48}n_f^2\zeta_3+\frac{950963}{4096}l_{\mu Q}\notag\\
 &+\frac{5967}{1024}l_{\mu Q}\zeta_3-\frac{72907}{3072}l_{\mu Q}n_f\-\frac{51}{64}l_{\mu Q}n_f\zeta_3+\frac{263}{576}l_{\mu Q}n_f^2+\frac{61797}{1024}l_{\mu Q}^2\notag\\
 &-\frac{757}{128}l_{\mu Q}^2n_f+\frac{25}{192}l_{\mu Q}^2n_f^2+\frac{4929}{1024}l_{\mu Q}^3-\frac{115}{256}l_{\mu Q}^3n_f+\frac{1}{96}l_{\mu Q}^3n_f^2\bigg)\bigg]\,.
\end{align}
Here and everywhere in the paper the renormalization is carried out
in the $\msbar$-scheme, $n_f$ is the total number of quark flavours,
$l_{\mu Q}
= \ln \frac{\mu^2}{Q^2}$, $m = m(\mu)$ and $\as=
\frac{\alpha_s(\mu)}{\pi}$ are the running quark mass and quark-gluon
coupling constant respectively. In addition, the irrational constants
$
\zeta_3 = 1.2020569,\,\zeta_4 =   1.0823232,\, \zeta_5= 1.0369277$ appear. 
In numerical form  Eqs. \eqref{eq:c_g_m2}-\eqref{eq:c_h_A2}  read 
\begin{align}
  \label{eq:c_g_h_num}
C_{m^2}^g=&  a_s\big[1+a_s(17.7614-0.555556\,n_f)\notag\\
&+a_s^2(311.276-25.8972\,n_f+0.231481\,n_f^2)\big]\,,\\
C_{A^2}^g=&\frac{3}{8}\,\pi^2 \,
a_s\big[1+a_s(8.17708-0.611111\,n_f)\notag\\
&+a_s^2(86.9600-13.6255\,n_f+0.373457\,n_f^2)\notag\\
&+a_s^3(1068.69  -240.803\,n_f+14.1703\,n_f^2-0.191187\,n_f^3)\big]\,,\\
C_{m^2}^h=& -\frac{3}{8}a_s^2\big[1+a_s(33.2934-0.958333\,n_f)\big]\,,\\
C_{A^2}^h=&\frac{3}{8}\,\pi^2 \,
a_s\big[1+3.75000a_s\notag\\
&+a_s^2(30.0240-1.46094\,n_f)\notag\\
&+a_s^3(339.141-38.2844\,n_f+0.655878\,n_f^2)\big]\,,
\end{align}
where we have set $\mu = Q$.

\section{OPE for the quark propagator}

The quark propagator of a quark field $\psi_q$ with mass $m$ 
is expressed in terms of the corresponding dressing functions as follows:
\begin{equation}
\mathrm{i} \int \mathrm{d}x \, \mathrm{e}^{\mathrm{i}qx}
\langle \mathrm{T}[\psi (x) \bar\psi  (0)] \rangle
= 
\frac{\FMslash{\! q}}{Q^2}\, V(Q) + \frac{{\displaystyle{S(Q)} }}{Q^2}
\label{quark_prop}
\end{equation}
The OPE expansions for the dressing functions (up to operators of dimension three) are
\begin{equation}
V(Q) = V_0(\mu/Q,\as) + \frac{C^q_{m^2}(\mu/Q,\as)}{Q^2} \, m^2
+ \frac{C^q_{A^2} (\mu/Q,\as)}{Q^2}\langle A^2\rangle
\label{OPE:V}
{},
\end{equation}

\begin{equation}
S(Q) = S_0(\mu/Q,\as)\, m + \frac{C^q_{m^3}(\mu/Q,\as)}{Q^2} \, m^3
+
\frac{ C^q_{A^2} (\mu/Q,\as)}{Q^2}\langle m A^2\rangle
+ \frac{C^q_{\bar{\psi}\psi}(\mu/Q,\as)}{Q^2}\langle \bar{\psi}\psi\rangle
\label{OPE:S}
{},
\end{equation}
where $m$ is the quark mass of the quark associated with the quark
field $\psi$.

\ice{
that is,  if $\psi$ is considered as massless one should
set to zero all $m$-dependent contributions to (\ref{OPE:V},\ref{OPE:S})
}

The purely perturbative  contributions $V_0$ and $S_0$ have been already discussed at
three-loop level in \cite{Chetyrkin:1999pq},  the files with results in computer readable form
can be downloaded from \verb+http://www-ttp.particle.uni-karlsruhe.de/Progdata/ttp99/ttp99-43+.
The remaining coefficient functions are listed below:
\begin{align}
\label{eq:c_q_m2}
C_{m^2}^q=&-\bigg[1+a_s\bigg(\frac{8}{3}+2l_{\mu Q}\bigg)+a_s^2\bigg(\frac{617}{24}-\frac{10}{3}\zeta_3-\frac{121}{144}n_f\notag\\
 &+\frac{307}{16}l_{\mu Q}-\frac{23}{36}l_{\mu Q}n_f+\frac{19}{4}l_{\mu Q}^2-\frac{1}{6}l_{\mu Q}^2n_f\bigg)\notag\\
 &+a_s^3\bigg(\frac{58211}{192}-\frac{17}{256}\zeta_4+\frac{2165}{144}\zeta_5-\frac{279733}{3456}\zeta_3-\frac{173449}{7776}n_f\notag\\
 &-\frac{5}{6}n_f\zeta_4+\frac{625}{216}n_f\zeta_3+\frac{2999}{23328}n_f^2+\frac{1}{27}n_f^2\zeta_3+\frac{64803}{256}l_{\mu Q}\notag\\
 &-\frac{3217}{128}l_{\mu Q}\zeta_3-\frac{685}{36}l_{\mu Q}n_f-\frac{5}{9}l_{\mu Q}n_f\zeta_3+\frac{139}{648}l_{\mu Q}n_f^2+\frac{2045}{24}l_{\mu Q}^2\notag\\
 &-\frac{1895}{288}l_{\mu Q}^2n_f+\frac{23}{216}l_{\mu Q}^2n_f^2+\frac{95}{8}l_{\mu Q}^3-\frac{17}{18}l_{\mu Q}^3n_f+\frac{1}{54}l_{\mu Q}^3n_f^2\bigg)\displaybreak[0]\\
\label{eq:c_q_A2}
 C_{A^2}^q=&-\frac{1}{3}\pi^2a_s\bigg[1+a_s\bigg(\frac{3}{4}+\frac{9}{16}l_{\mu Q}\bigg)+a_s^2\bigg(\frac{32167}{9216}-\frac{117}{128}\zeta_3\notag\\
 &-\frac{137}{768}n_f+\frac{1691}{768}l_{\mu Q}-\frac{23}{192}l_{\mu Q}n_f+\frac{477}{512}l_{\mu Q}^2-\frac{3}{64}l_{\mu Q}^2n_f\bigg)\notag\\
 &+a_s^3\bigg(\frac{13735835}{663552}+\frac{1507}{4096}\zeta_4-\frac{11205}{2048}\zeta_5+\frac{16481}{4096}\zeta_3-\frac{207901}{82944}n_f\notag\\
 &-\frac{33}{256}n_f\zeta_4+\frac{419}{1152}n_f\zeta_3+\frac{3139}{124416}n_f^2+\frac{1}{96}n_f^2\zeta_3+\frac{3036731}{147456}l_{\mu Q}\notag\\
 &-\frac{4921}{1024}l_{\mu Q}\zeta_3-\frac{253765}{110592}l_{\mu Q}n_f+\frac{3}{64}l_{\mu Q}n_f\zeta_3+\frac{113}{3456}l_{\mu Q}n_f^2+\frac{103069}{12288}l_{\mu Q}^2\notag\\
 &-\frac{8767}{9216}l_{\mu Q}^2n_f+\frac{23}{1152}l_{\mu Q}^2n_f^2+\frac{15423}{8192}l_{\mu Q}^3-\frac{203}{1024}l_{\mu Q}^3n_f+\frac{1}{192}l_{\mu Q}^3n_f^2\bigg)\displaybreak[0]\\
\label{eq:c_q_m3}
 C_{m^3}^q=&-\bigg[1+a_s(4+2l_{\mu Q})\notag\\
 &+a_s^2\bigg(\frac{3545}{96}-\frac{2}{3}\zeta_3-\frac{5}{4}n_f+\frac{641}{24}l_{\mu Q}-\frac{13}{18}l_{\mu Q}n_f+\frac{39}{8}l_{\mu Q}^2-\frac{1}{12}l_{\mu Q}^2n_f\bigg)\notag\\
 &+a_s^3\bigg(\frac{9287323}{20736}+\frac{493}{768}\zeta_4+\frac{1975}{54}\zeta_5-\frac{63643}{864}\zeta_3-\frac{523}{16}n_f\notag\\
 &-\frac{5}{4}n_f\zeta_4-\frac{55}{216}n_f\zeta_3+\frac{383}{1944}n_f^2+\frac{1}{6}n_f^2\zeta_3+\frac{424327}{1152}l_{\mu Q}\notag\\
 &-\frac{241}{64}l_{\mu Q}\zeta_3-\frac{10375}{432}l_{\mu Q}n_f-\frac{13}{9}l_{\mu Q}n_f\zeta_3+\frac{20}{81}l_{\mu Q}n_f^2\notag\\
 &+\frac{7401}{64}l_{\mu Q}^2-\frac{211}{32}l_{\mu Q}^2n_f+\frac{2}{27}l_{\mu Q}^2n_f^2+\frac{25}{2}l_{\mu Q}^3-\frac{5}{9}l_{\mu Q}^3n_f\bigg)\displaybreak[0]\\
\label{eq:c_q_mA2}
 C_{mA^2}^q=&\frac{25}{48}\pi^2a_s^2\bigg[1+a_s\bigg(\frac{4409}{400}-\frac{373}{900}n_f+\frac{69}{16}l_{\mu Q}-\frac{1}{6}l_{\mu Q}n_f\bigg)\notag\\
 &+a_s^2\bigg(\frac{35490283}{230400}-\frac{72037}{4800}\zeta_3-\frac{2219557}{172800}n_f+\frac{29}{900}n_f\zeta_3\notag\\
 &+\frac{3011}{16200}n_f^2+\frac{282071}{3200}l_{\mu Q}-\frac{54191}{7200}l_{\mu Q}n_f+\frac{373}{2700}l_{\mu Q}n_f^2\notag\\
 &+\frac{7797}{512}l_{\mu Q}^2-\frac{251}{192}l_{\mu Q}^2n_f+\frac{1}{36}l_{\mu Q}^2n_f^2\bigg)\displaybreak[0]\\
\label{eq:c_q_qq}
 C_{\bar{\psi}\psi}^q=&-\frac{4}{3}\pi^2a_s\bigg[1+a_s\bigg(\frac{99}{16}-\frac{5}{18}n_f+\frac{7}{4}l_{\mu Q}-\frac{1}{6}l_{\mu Q}n_f\bigg)\notag\\
 &+a_s^2\bigg(\frac{13745}{256}-\frac{79}{128}\zeta_3-\frac{1193}{216}n_f-\frac{5}{6}n_f\zeta_3+\frac{25}{324}n_f^2+\frac{2747}{96}l_{\mu Q}\notag\\
 &-\frac{559}{144}l_{\mu Q}n_f+\frac{5}{54}l_{\mu Q}n_f^2+\frac{63}{16}l_{\mu Q}^2-\frac{2}{3}l_{\mu Q}^2n_f+\frac{1}{36}l_{\mu Q}^2n_f^2\bigg)\notag\\
 &+a_s^3\bigg(\frac{26331733}{41472}+\frac{79}{256}\zeta_4-\frac{12166325}{331776}\zeta_5-\frac{2236285}{82944}\zeta_3\notag\\
 &-\frac{403157}{3888}n_f+\frac{5}{12}n_f\zeta_4+\frac{70}{9}n_f\zeta_5-\frac{8209}{576}n_f\zeta_3\notag\\
 &+\frac{722269}{186624}n_f^2+\frac{301}{432}n_f^2\zeta_3-\frac{125}{5832}n_f^3+\frac{3937861}{9216}l_{\mu Q}-\frac{1975}{512}l_{\mu Q}\zeta_3\notag\\
 &-\frac{1097999}{13824}l_{\mu Q}n_f-\frac{3763}{768}l_{\mu Q}n_f\zeta_3+\frac{39487}{10368}l_{\mu Q}n_f^2+\frac{5}{12}l_{\mu Q}n_f^2\zeta_3\notag\\
 &-\frac{25}{648}l_{\mu Q}n_f^3+\frac{28155}{256}l_{\mu Q}^2-\frac{13255}{576}l_{\mu Q}^2n_f+\frac{2453}{1728}l_{\mu Q}^2n_f^2\notag\\
 &-\frac{5}{216}l_{\mu Q}^2n_f^3+\frac{609}{64}l_{\mu Q}^3-\frac{653}{288}l_{\mu Q}^3n_f+\frac{77}{432}l_{\mu Q}^3n_f^2-\frac{1}{216}l_{\mu Q}^3n_f^3\bigg)
\end{align}
Their numerical form (with $\mu=Q$) reads:
\begin{align}
  \label{eq:c_q_num}
C_{m^2}^q=& -  \big[1+2.66667a_s+a_s^2(21.7015-0.840278\,n_f)\notag\\
 &+a_s^3(221.404-19.7294\,n_f+0.173079\,n_f^2)\big]\\
C_{A^2}^q=& -\frac{\pi^2}{3}  a_s
\big[1+0.750000a_s+a_s^2(2.39159-0.178385\,n_f)\notag\\
 &+a_s^3(20.2621-2.20883\,n_f+0.0377513\,n_f^2)\big]\\
C_{m^3}^q=& -  \big[1+4.00000a_s+a_s^2(36.1257-1.25000\,n_f)\notag\\
 &+a_s^3(397.959-34.3465\,n_f+0.397359\,n_f^2)\big]\\
C_{mA^2}^q=& \frac{25\pi^2}{48}  a_s^2
\big[1+a_s(11.0225-0.414444\,n_f)\notag\\
&+a_s^2(135.998-12.8059\,n_f+0.185864\,n_f^2)\big]\\
C_{\bar{\psi}\psi}^q=& -\frac{4\pi^2}{3}   a_s
\big[1+a_s(6.18750-0.277778\,n_f)\notag\\
 &+a_s^2(52.9495-6.52486\,n_f+0.0771605\,n_f^2)\notag\\
 &+a_s^3(564.828-112.308\,n_f+4.70773\,n_f^2-0.0214335\,n_f^3)\big]
\end{align}

\section{Renormalization Group Improvement}

\subsection{Anomalous dimensions}
\label{anom_dim}

We limit ourselves to the three-loop level.
Let us start from the operators of dimension two. The corresponding 
matrix of anomalous dimensions is defined by the following matrix equation 
\begin{equation}
  \label{eq:gam_dim2}
  \mu^2 \frac{\text{d}}{\text{d}\mu^2} 
\left(\begin{array}{c}
      A^2\\
      m^2
    \end{array}\right) =
\left(\begin{array}{cc}
    \gamma_{A^2} & \gamma_{A^2,m^2}\\
    0           & 2\gamma_m
  \end{array}\right)
\left(\begin{array}{c}
    A^2 \\
    m^2
  \end{array}\right)\,,
\end{equation}
where the differentiation on the lhs is carried out with fixed bare
coupling and quark masses.  The quark mass anomalous dimension is known
since long \cite{Tarasov:1982gk,Larin:1993tp} and the
anomalous dimension of $A^2$ in Landau gauge was found in \cite{Gracey:2002yt} to be
\begin{equation}
  \label{eq:gamma_A2}
  \begin{split}
    \gamma_{A^2}=\,  
&a_s\left(\frac{35}{16}-\frac{n_f}{6}\right)+a_s^2\left(\frac{1347}{256}-\frac{137n_f}{192}\right)\\
 &+a_s^3\left(\frac{75607}{4096}-\frac{18221n_f}{4608}+\frac{755n_f^2}{6912}-\frac{243\zeta_3}{2048}+\frac{33n_f\zeta_3}{128}\right)\\
 &+a_s^4\left(\frac{29764511}{393216}-\frac{57858155n_f}{2654208}+\frac{46549n_f^2}{41472}+\frac{6613n_f^3}{746496}\right.\\
 &\quad-\frac{99639\zeta_3}{131072}+\frac{335585n_f\zeta_3}{110592}+\frac{8489n_f^2\zeta_3}{41472}-\frac{n_f^3\zeta_3}{192}+\frac{8019\zeta_4}{16384}\\
 &\quad\left.-\frac{8955n_f\zeta_4}{8192}+\frac{33n_f^2\zeta_4}{512}+\frac{40905\zeta_5}{2048}-\frac{3355n_f\zeta_5}{1024}\right)
{}.
  \end{split}
\end{equation}

The non-diagonal  three-loop
anomalous dimension $\gamma_{A^2,m^2}$ reads
\begin{equation}
  \label{eq:gam_A2m2}
\gamma_{A^2,m^2}=\frac{a_s}{16\pi^2}\bigg[24+a_s\bigg(\frac{971}{4}-4\,n_f+36\,\zeta_3\bigg)
\bigg]
{}.
\end{equation}
Life is easier with operators of dimension three.  
First, the anomalous dimensions of the pair 
$m\,A^2$ and $m^3$ are, obviously, additively
related to  those considered above, namely:
\begin{equation}
  \label{eq:gam_mx}
\gamma_{m\,A^2} = \gamma_m +\gamma_{A^2}\,, \qquad \gamma_{m^3} = 3\gamma_m\,.
\end{equation}
Second, in the process of renormalization  the quark condensate  could mix  only 
with the unit  operator (times a   quark mass cubed):
\begin{equation}
  \label{eq:gam_dim3}
  \mu^2 \frac{\text{d}}{\text{d}\mu^2} 
\left(\begin{array}{c}
      \bar{\psi}\psi\\
      m^3
    \end{array}\right) =
\left(\begin{array}{cc}
    \gamma_{\bar{\psi}\psi} & \gamma_{\bar{\psi}\psi,m^3}\\
    0           & 3\gamma_m
  \end{array}\right)
\left(\begin{array}{c}
    \bar{\psi}\psi \\
    m^3
  \end{array}\right)\,.  
\end{equation}
The fact that 
\begin{equation}
\gamma_{\bar{\psi}\psi} \equiv -\gamma_m 
\end{equation}
is well-known from text-books.  The non-diagonal part of the mixing
was investigated in detail a long time ago \cite{Spiridonov:1988md,Chetyrkin:1994qu}. It is naturally
expressed in terms of the so-called vacuum anomalous dimension, $\gamma^d_0$
as follows \cite{Chetyrkin:1994ex}:
\begin{equation}
\mu^2\frac{\mathrm{d}}{d \mu^2} {\bar{\psi}\psi} = -\gamma_m \, {\bar{\psi}\psi} - 
4 m^3 \, \gamma^d_0(\as)
{},
\end{equation}
with
\[
\gamma^d_0 =  -\frac{3}{16\pi^2}
\left[  1 + \frac{4}{3} a_s
 + \left(\frac{313}{72}- \frac{5}{12} n_f - \frac{2}{3}\zeta_3 \right)a_s^2
\right]
{}.
\]
 
\subsection{Scheme-independent correlators and  operators}

In general a Green function $G$ depends on the renormalization prescription
({\it scheme}) and the choice of the artificial scale $\mu$. It is,
however, well-known how  to define a variant $\hat{G}$ of $G$ which is invariant
under changes of the renormalization scheme and $\mu$. 
The corresponding renormalization group equation (RGE) 
\begin{equation}
  \label{eq:sc_inv_def}
  \mu^2 \frac{\text{d}}{\text{d}\mu^2}\hat{G} = 0
\end{equation}
has the formal solution
\begin{equation}
  \label{eq:sc_inv_sol}
  \hat{G} = G(a_s,\mu)/f(a_s),\qquad f(a_s) =
  \exp\left(\int\frac{\text{d}a_s}{a_s}
    \frac{\gamma_G}{\beta}\right)\,.
\end{equation}
The OPE of a (suitable) scale-invariant Green function can be
rewritten in terms of scale-invariant operators $\hat{{\cal O}}_i$ and Wilson
coefficients $\hat{C}_i$, which again obey RGEs of the form
\eqref{eq:sc_inv_def}.

Here, we consider the OPEs of the scheme-independent dressing functions in
the limit of massless quarks. In this limit --- aside from the
perturbative contributions --- only the operators  $A^2$
and $\bar{\psi}\psi$ contribute. For the operator $A^2$
and its coefficient functions we obtain (in the three-loop approximation)
\begin{align}
  \label{eq:sc_inv_op:1}
  \widehat{A^2}\big|_{n_f=0} =&\, a_s^{-\frac{35}{44}}(1+0.0693440\,a_s+0.0240863\,a_s^2+0.405494\,a_s^3)\,A^2\,,\\  \label{eq:sc_inv_op:2}
  \widehat{A^2}\big|_{n_f=2} =&\, a_s^{-\frac{89}{116}}(1+0.0654912\,a_s+0.0933818\,a_s^2+0.508904\,a_s^3)\,A^2\,,\\  \label{eq:sc_inv_op:3}
  \widehat{A^2}\big|_{n_f=3} =&\, a_s^{-\frac{3}{4}}(1+0.0538194\,a_s+0.136131\,a_s^2+0.570436\,a_s^3)\,A^2\,,\\  \label{eq:sc_inv_op:4}
  \hat{C}_{A^2}^g\big|_{n_f=0}=&\,\frac{3}{8}\,\pi^2 a_s^{\,\frac{9}{44}}(a_s+8.24643\,a_s^2+87.5512\,a_s^3+1075.32\,a_s^4)\,,\\  \label{eq:sc_inv_op:5}
  \hat{C}_{A^2}^g\big|_{n_f=2}=&\,\frac{3}{8}\,\pi^2 a_s^{\,\frac{27}{116}}\left(a_s+7.02035\,a_s^2+61.7518\,a_s^3+647.400\,a_s^4\right)\,,\\  \label{eq:sc_inv_op:6}
  \hat{C}_{A^2}^g\big|_{n_f=3}=&\,\frac{3}{8}\,\pi^2 a_s^{\,\frac{1}{4}}(a_s+6.39757\,a_s^2+49.9224\,a_s^3+472.744\,a_s^4)\,,\\  \label{eq:sc_inv_op:7}
  \hat{C}_{A^2}^h\big|_{n_f=0}=&\,\frac{3}{8}\,\pi^2 a_s^{\,\frac{13}{22}}(a_s+3.61131\,a_s^2+29.4702\,a_s^3+334.048\,a_s^4)\,,\\  \label{eq:sc_inv_op:9}
  \hat{C}_{A^2}^h\big|_{n_f=2}=&\,\frac{3}{8}\,\pi^2 a_s^{\,\frac{31}{58}}\left(a_s+3.61902\,a_s^2+26.4370\,a_s^3+260.012\,a_s^4\right)\,,\\  \label{eq:sc_inv_op:10}
  \hat{C}_{A^2}^h\big|_{n_f=3}=&\,\frac{3}{8}\,\pi^2 \,\sqrt{a_s}(a_s+3.64236\,a_s^2+24.9740\,a_s^3+225.345\,a_s^4)\,,\\  \label{eq:sc_inv_op:11}
  \hat{C}_{A^2}^q\big|_{n_f=0}=&-\frac{\pi^2}{3}a_s^{\,\frac{35}{44}}(a_s+0.173080\,a_s^2+1.19104\,a_s^3+15.9766\,a_s^4)\,,\\  \label{eq:sc_inv_op:13}
  \hat{C}_{A^2}^q\big|_{n_f=2}=&-\frac{\pi^2}{3}a_s^{\,\frac{89}{116}}\left(a_s+0.175888\,a_s^2+0.859625\,a_s^3+12.6022\,a_s^4\right)\,,\\  \label{eq:sc_inv_op:14}
  \hat{C}_{A^2}^q\big|_{n_f=3}=&-\frac{\pi^2}{3}a_s^{\,\frac{3}{4}}(a_s+0.186921\,a_s^2+0.689610\,a_s^3+11.0348\,a_s^4)
\end{align}
by inserting the corresponding anomalous dimensions into
Eq. \eqref{eq:sc_inv_sol}. The scale- and scheme-independent versions of
$\bar{\psi}\psi$ and its coefficient function in the OPE of the quark
propagator read
\begin{align}
  \label{eq:sc_inv_c:1}
  \widehat{\bar{\psi}\psi}\big|_{n_f=0} =&\, a_s^{-\frac{4}{11}}(1+0.687328\,a_s+1.51211\,a_s^2+4.05787\,a_s^3)\,\bar{\psi}\psi\,,\\ \label{eq:sc_inv_c:2}
  \widehat{\bar{\psi}\psi}\big|_{n_f=2} =&\,a_s^{-\frac{12}{29}}(1+0.805985\,a_s+1.40095\,a_s^2+2.72916\,a_s^3)\, \bar{\psi}\psi\,,\\ \label{eq:sc_inv_c:3}
  \widehat{\bar{\psi}\psi}\big|_{n_f=3} =&\, a_s^{-\frac{4}{9}}(1+0.895062\,a_s+1.37143\,a_s^2+1.95168\,a_s^3)\,\bar{\psi}\psi\,,\\ \label{eq:sc_inv_c:4}
  \hat{C}_{\bar{\psi}\psi}^q\big|_{n_f=0}=&\,-\frac{4\pi^2}{3}a_s^{\,\frac{4}{11}}(a_s+4.99260\,a_s^2+44.0815\,a_s^3+489.206\,a_s^4)\,,\\ \label{eq:sc_inv_c:5}
  \hat{C}_{\bar{\psi}\psi}^q\big|_{n_f=2}=&\,-\frac{4\pi^2}{3}a_s^{\,\frac{12}{29}}\left(a_s+4.31734\,a_s^2+31.7744\,a_s^3+298.894\,a_s^4\right)\,,\\ \label{eq:sc_inv_c:6}
  \hat{C}_{\bar{\psi}\psi}^q\big|_{n_f=3}=&\,-\frac{4\pi^2}{3}a_s^{\,\frac{4}{9}}(a_s+3.94985\,a_s^2+25.7972\,a_s^3+217.583\,a_s^4)\,.
\end{align}


Another useful scheme-invariant object is the so-called  ``effective quark mass''  $m_P(Q)$ which is defined as follows 
\cite{Politzer:1976tv}.
\begin{equation}
\mathrm{i} \int \mathrm{d}x \, \mathrm{e}^{\mathrm{i}qx}
\langle \mathrm{T}[\psi (x) \bar\psi  (0)] \rangle
= \frac{1}{B - A\,\FMslash{\! q}}
{},
 \ \ \  m_P(Q) = \frac{B(q)}{A(q)}  = \frac{S(Q)}{V(Q)}
\label{quark_prop2}
{},
\end{equation}
where we have used Eq.~\re{quark_prop2} to express  $m_P(Q)$ in terms of 
the dressing functions $V(Q)$ and $S(Q)$.

In the chiral limit the leading contribution to  $m_P(Q)$  comes from
the quark condensate; in explicit form we get
\begin{equation}
m_P(Q) = C_{\bar{\psi}\psi}^q
V_0^{-1} \frac{\langle \bar{\psi}\psi \rangle}{Q^2}\,.
\label{eq:mp}
\end{equation}
 Using 
the results of Section 3 we arrive at:
\begin{align}
m_P(Q)|_{n_f=0} =&\,-\frac{4\pi^2}{3}a_s \bigg[
1+a_s\bigg(\frac{99}{16}\bigg)+a_s^2\bigg(\frac{129449}{2304}-\frac{175}{128}\zeta_3\bigg)\\
 &+a_s^3\bigg(\frac{28729643}{41472}-\frac{10153205}{331776}\zeta_5-\frac{4351141}{82944}\zeta_3\bigg)\bigg]\frac{\langle \bar{\psi}\psi \rangle}{Q^2}\,,
\label{eq:mP_0}
\\
m_P(Q)|_{n_f=1} =&\,-\frac{4\pi^2}{3}a_s\bigg[1+a_s\bigg(\frac{851}{144}\bigg)+a_s^2\bigg(\frac{1049089}{20736}-\frac{845}{384}\zeta_3\bigg)\\
&+a_s^3\bigg(\frac{72992597}{124416} -\frac{7572725}{331776}\zeta_5+\frac{5}{12}\zeta_4-\frac{5394805}{82944}\zeta_3\bigg)\bigg]\frac{\langle \bar{\psi}\psi \rangle}{Q^2}\,,
\label{eq:mP_1}
\\
m_P(Q)|_{n_f=2} =&\,-\frac{4\pi^2}{3}a_s\bigg[1+a_s\bigg(\frac{811}{144}\bigg)+a_s^2\bigg(\frac{936337}{20736}-\frac{1165}{384}\zeta_3\bigg)\\
 &+a_s^3\bigg(\frac{182335471}{373248}-\frac{4992245}{331776}\zeta_5+\frac{5}{6}\zeta_4-\frac{6322885}{82944}\zeta_3\bigg)\bigg]\frac{\langle \bar{\psi}\psi \rangle}{Q^2}\,,
\label{eq:mP_2}
\\
m_P(Q)|_{n_f=3} =&\,-\frac{4\pi^2}{3}a_s\bigg[1+a_s\bigg(\frac{257}{48}\bigg)+a_s^2\bigg(\frac{91865}{2304}-\frac{495}{128}\zeta_3\bigg)\\
 &+a_s^3\bigg(\frac{611489}{1536}-\frac{2411765}{331776}\zeta_5+\frac{5}{4}\zeta_4-\frac{7135381}{82944}\zeta_3\bigg)\bigg]\frac{\langle \bar{\psi}\psi \rangle}{Q^2}
\label{eq:mP_3}\,,
\end{align}
where we have set the  renormalization scale $\mu =Q$.
Numerically these equations read: 
\begin{align}
  m_P(Q)|_{n_f=0} =&\,-\frac{4\pi^2}{3}a_s(1 + 6.1875 \,a_s + 54.541 \,a_s^2 + 597.957 \,a_s^3)\frac{\langle \bar{\psi}\psi \rangle}{Q^2}\,,
  \label{eq:mP_num_0}
  \\
  m_P(Q)|_{n_f=1} =&\,-\frac{4\pi^2}{3}a_s(1 + 5.90972 \,a_s + 47.9475 \,a_s^2 + 485.281 \,a_s^3)\frac{\langle \bar{\psi}\psi \rangle}{Q^2}\,,
  \label{eq:mP_num_1}
  \\
  m_P(Q)|_{n_f=2} =&\,-\frac{4\pi^2}{3}a_s(1 + 5.63194 \,a_s + 41.5083 \,a_s^2 + 382.176 \,a_s^3)\frac{\langle \bar{\psi}\psi \rangle}{Q^2}\,,
  \label{eq:mP_num_2}
  \\
  m_P(Q)|_{n_f=3} =&\,-\frac{4\pi^2}{3}a_s(1 + 5.35417 \,a_s + 35.2234 \,a_s^2 + 288.511 \,a_s^3)\frac{\langle \bar{\psi}\psi \rangle}{Q^2}\,.
  \label{eq:mP_num_3}
\end{align}

In analogy to the effective quark mass one can also define effective
masses for gluon and ghost fields, which are induced by the gluon mass
condensate $\langle A^2 \rangle$. An explicit formula can be derived by
considering the ghost propagator (or the gluon propagator) in the chiral limit:
\begin{equation}
  \label{eq:ghost_massive_prop}
  \Delta^{ab}(q) = \frac{\delta^{ab}}{Q^2}
\ \
 \left(D_0^h(Q)+C_{A^2}^h(Q)\frac{\langle A^2 \rangle}{Q^2}\right)
 \approx
\frac{\delta^{ab}}{Q^2-C_{A^2}^h(Q)/D_0^h(Q)\langle A^2 \rangle}
\ \
 D_0^h(Q)\,.
\end{equation}
The effective masses are then given by
\begin{equation}
  \label{eq:g_h_mass}
  m^2_?(Q) = - \frac{C_{A^2}^?(Q)}{D_0^?(Q)}\langle A^2 \rangle\,,
\end{equation}
where $?$ stands for $g$ or $h$. The analytic results for
 $\mu=Q$ and $n_f=1,\,2,\,3$ are 
\begin{align}
  \label{eq:mg_0}
m^2_g|_{n_f=0}=&-\frac{3}{8}\pi^2 a_s\bigg[1+a_s\bigg(\frac{197}{32}\bigg)+a_s^2\bigg(\frac{480587}{9216}+\frac{243}{128}\zeta_3\bigg)\notag\\
 &+a_s^3\bigg(\frac{520248245}{884736}-\frac{82215}{4096}\zeta_5+\frac{243}{4096}\zeta_4+\frac{331077}{8192}\zeta_3\bigg)\notag\\
 &+a_s^4\bigg(-\frac{38766561211}{7077888}+\frac{25948995}{131072}\zeta_5\notag\\
 &+\frac{35721}{32768}\zeta_4+\frac{57028439}{131072}\zeta_3+\frac{729}{2048}\zeta_3^2\bigg)\bigg]\langle A^2 \rangle \,,\displaybreak[0]\\
  \label{eq:mg_1}
m^2_g|_{n_f=1}=&-\frac{3}{8}\pi^2 a_s\bigg[1+a_s\bigg(\frac{559}{96}\bigg)+a_s^2\bigg(\frac{3836375}{82944}+\frac{195}{128}\zeta_3\bigg)\notag\\
 &+a_s^3\bigg(\frac{11737602763}{23887872}-\frac{74855}{4096}\zeta_5-\frac{285}{4096}\zeta_4+\frac{7127207}{221184}\zeta_3\bigg)\notag\\
 &+a_s^4\bigg(-\frac{2182125807293}{573308928}+\frac{1465676035}{10616832}\zeta_5\notag\\
 &-\frac{38475}{32768}\zeta_4+\frac{12243145349}{31850496}\zeta_3-\frac{8075}{18432}\zeta_3^2\bigg)\bigg]\langle A^2 \rangle \,,\displaybreak[0]\\
  \label{eq:mg_2}
m^2_g|_{n_f=2}=&-\frac{3}{8}\pi^2 a_s\bigg[1+a_s\bigg(\frac{527}{96}\bigg)+a_s^2\bigg(\frac{3368459}{82944}+\frac{147}{128}\zeta_3\bigg)\notag\\
 &+a_s^3\bigg(\frac{9610131071}{23887872}-\frac{67495}{4096}\zeta_5-\frac{813}{4096}\zeta_4+\frac{5402375}{221184}\zeta_3\bigg)\notag\\
 &+a_s^4\bigg(-\frac{477599196341}{191102976}+\frac{916103635}{10616832}\zeta_5\notag\\
 &-\frac{99999}{32768}\zeta_4+\frac{3506053415}{10616832}\zeta_3-\frac{24119}{18432}\zeta_3^2\bigg)\bigg]\langle A^2 \rangle \,,\displaybreak[0]\\
  \label{eq:mg_3}
m^2_g|_{n_f=3}=&-\frac{3}{8}\pi^2 a_s\bigg[1+a_s\bigg(\frac{165}{32}\bigg)+a_s^2\bigg(\frac{108205}{3072}+\frac{99}{128}\zeta_3\bigg)\notag\\
 &+a_s^3\bigg(\frac{31529153}{98304}-\frac{60135}{4096}\zeta_5-\frac{1341}{4096}\zeta_4+\frac{139429}{8192}\zeta_3\bigg)\notag\\
 &+a_s^4\bigg(-\frac{10684895233}{7077888}+\frac{16783385}{393216}\zeta_5-\notag\\
 &\frac{148851}{32768}\zeta_4+\frac{322973567}{1179648}\zeta_3-\frac{4619}{2048}\zeta_3^2\bigg)\bigg]\langle A^2 \rangle \,,\displaybreak[0]\\
  \label{eq:mh_0}
 m^2_h|_{n_f=0}=&-\frac{3}{8}\pi^2 a_s\bigg[1+3\,a_s+a_s^2\bigg(\frac{20997}{1024}+\frac{351}{256}\zeta_3\bigg)\notag\\
 &+a_s^3\bigg(\frac{4345483}{18432}-\frac{53235}{4096}\zeta_5+\frac{243}{4096}\zeta_4+\frac{24945}{1024}\zeta_3\bigg)\notag\\
 &+a_s^4\bigg(-\frac{1004316139}{1572864}+\frac{222885}{16384}\zeta_5-\frac{5103}{16384}\zeta_4+\frac{4937751}{131072}\zeta_3+\frac{3645}{8192}\zeta_3^2\bigg)\bigg]\langle A^2 \rangle \,,\displaybreak[0]\\
  \label{eq:mh_1}
 m^2_h|_{n_f=1}=&-\frac{3}{8}\pi^2 a_s\bigg[1+3\,a_s+a_s^2\bigg(\frac{19881}{1024}+\frac{351}{256}\zeta_3\bigg)\notag\\
 &+a_s^3\bigg(\frac{34790023}{165888}-\frac{53235}{4096}\zeta_5-\frac{285}{4096}\zeta_4+\frac{69767}{3072}\zeta_3\bigg)\notag\\
 &+a_s^4\bigg(-\frac{8001655387}{14155776}+\frac{222885}{16384}\zeta_5+\frac{5985}{16384}\zeta_4+\frac{4761971}{131072}\zeta_3+\frac{3645}{8192}\zeta_3^2\bigg)\bigg]\langle A^2 \rangle \,,\displaybreak[0]\\
  \label{eq:mh_2}
 m^2_h=|_{n_f=2}&-\frac{3}{8}\pi^2 a_s\bigg[1+3\,a_s+a_s^2\bigg(\frac{18765}{1024}+\frac{351}{256}\zeta_3\bigg)\notag\\
 &+a_s^3\bigg(\frac{30638707}{165888}-\frac{53235}{4096}\zeta_5-\frac{813}{4096}\zeta_4+\frac{64763}{3072}\zeta_3\bigg)\notag\\
 &+a_s^4\bigg(-\frac{7006421587}{14155776}+\frac{222885}{16384}\zeta_5+\frac{17073}{16384}\zeta_4+\frac{4571855}{131072}\zeta_3+\frac{3645}{8192}\zeta_3^2\bigg)\bigg]\langle A^2 \rangle \,,\displaybreak[0]\\
  \label{eq:mh_3}
 m^2_h|_{n_f=3}=&-\frac{3}{8}\pi^2 a_s\bigg[1+3\,a_s+a_s^2\bigg(\frac{17649}{1024}+\frac{351}{256}\zeta_3\bigg)\notag\\
 &+a_s^3\bigg(\frac{329079}{2048}-\frac{53235}{4096}\zeta_5-\frac{1341}{4096}\zeta_4+\frac{19941}{1024}\zeta_3\bigg)\notag\\
 &+a_s^4\bigg(-\frac{224190513}{524288}+\frac{222885}{16384}\zeta_5+\frac{28161}{16384}\zeta_4+\frac{4367403}{131072}\zeta_3+\frac{3645}{8192}\zeta_3^2\bigg)\bigg]\langle A^2 \rangle \,.
\end{align}
In numerical form we obtain
\begin{align}
  \label{eq:mg_num_0}
m^2_g|_{n_f=0}=&-\frac{3}{8}\pi^2a_s[1 + 6.15625\,a_s + 54.4291\,a_s^2 + 615.858\,a_s^3 - 4747.15\,a_s^4]\langle A^2 \rangle \,,\\
  \label{eq:mg_num_1}
m^2_g|_{n_f=1}=&-\frac{3}{8}\pi^2a_s[1 + 5.82292\,a_s + 48.0839\,a_s^2 + 511.071\,a_s^3 - 3202.89\,a_s^4]\langle A^2 \rangle \,,\\
  \label{eq:mg_num_2}
m^2_g|_{n_f=2}=&-\frac{3}{8}\pi^2a_s[1 + 5.48958\,a_s + 41.9917\,a_s^2 + 414.36\,a_s^3 - 2017.93\,a_s^4]\langle A^2 \rangle \,,\\
  \label{eq:mg_num_3}
m^2_g|_{n_f=3}=&-\frac{3}{8}\pi^2a_s[1 + 5.15625\,a_s + 36.1527\,a_s^2 + 325.612\,a_s^3 - 1144.42\,a_s^4]\langle A^2 \rangle \,,\\
  \label{eq:mh_num_0}
m^2_h|_{n_f=0}=&-\frac{3}{8}\pi^2a_s[1 + 3\,a_s + 22.153\,a_s^2 + 251.628\,a_s^3 - 578.831\,a_s^4]\langle A^2 \rangle \,,\\
  \label{eq:mh_num_1}
m^2_h|_{n_f=1}=&-\frac{3}{8}\pi^2a_s[1 + 3\,a_s + 21.0632\,a_s^2 + 223.467\,a_s^3 - 506.441\,a_s^4]\langle A^2 \rangle \,,\\
  \label{eq:mh_num_2}
m^2_h|_{n_f=2}=&-\frac{3}{8}\pi^2a_s[1 + 3\,a_s + 19.9733\,a_s^2 + 196.345\,a_s^3 - 437.146\,a_s^4]\langle A^2 \rangle \,,\\
  \label{eq:mh_num_3}
m^2_h|_{n_f=3}=&-\frac{3}{8}\pi^2a_s[1 + 3\,a_s + 18.8835\,a_s^2 + 170.26\,a_s^3 - 370.947\,a_s^4]\langle A^2 \rangle \,.
\end{align}

\section{Theoretical and hardware  tools}

In our work we have heavily used the software packages QGRAF
\cite{Nogueira:1991ex},  EXP \cite{Seidensticker:1999bb} and a modern
version of MINCER \cite{Gorishnii:1989gt,Larin:1991fz}
written in the algebraic computer language \mbox{FORM 3} \cite{Vermaseren:2000nd}
 for the generation and calculation of the required diagrams. 

The calculations have have been performed in a general covariant gauge
and  for the gauge group SU(n). The total number of diagrams contributing to
different channels (according to QGRAF) are displayed in Table 1. The
files with all results (in a computer readable form) can be
downloaded from  \verb+http://www-ttp.particle.uni-karlsruhe.de/Progdata/ttp09/ttp09-40/+

Any coefficient function of any operator entering into  the OPE of two
local operators can be expressed in terms of massless propagators. 
The reduction to massless propagators is  conveniently done  with the well-known
method of projectors \cite{Gorishnii:1983su,Gorishnii:1986gn}. 

Finally, the package MINCER is able to compute 
very effectively massless propagators up to (and including)
three loop level. 
\ice{We have used the local cluster...
The calculation took (very roughly) a month of CPU time (normalized on
a one   working station with 4 CPU of 2 Gz). }

\begin{table}
  \centering
  \begin{tabular}{lcccc}
\hline                       & tree & one loop & two loops & three loops \\\hline
$C^g_{m^2}$            & 1    & 5        & 59        & 1148        \\
$C^g_{A^2}$            & 6    & 222      & 7407      & 264399      \\
$C^h_{m^2}$            & 1    & 1        & 9         & 148         \\
$C^h_{A^2}$            & 2    & 23       & 595       & 19419       \\
$C^q_{m^n}$            & 1    & 1        & 9         & 148         \\
$C^q_{m^nA^2}$         & 2    & 23       & 657       & 23251       \\
$C^q_{\bar{\psi}\psi}$ & 1    & 11       & 234       & 6641        \\\hline
  \end{tabular}
  \caption{Number of diagrams contributing to the various coefficient functions}
\end{table}
\section{Conclusion}

We have computed the coefficient functions of the operators of
dimension two and three in the OPE for the gluon, ghost and quark
propagators. The higher order corrections are essential  as one
could see by inspecting eqs. \re{eq:sc_inv_op:1} -\re{eq:sc_inv_c:6}.  They are most important in
two cases: gluon condensate contributions to ghost and gluon
propagators as well as the quark condensate one to the quark
propagator. In general the terms proportional to $n_f$ tend to
significantly stabilize the perturbative series by decreasing the value
of higher order terms (cmp. e.g. eq.\re{eq:sc_inv_op:4} and \re{eq:sc_inv_op:7}).

Specific numerical analysis should be made with  
specific lattice data. Still,  we observe that the   higher order corrections
to the coeffcient functions  display relatively  good (apparent) convergency pattern
with basically positive coefficients which, presumably, should lead  
to a noticeable decrease of the value of the  $A^2$  condensate once the
lattice data are reanalyzed  with an account of newly computed terms in the 
corresponding OPE. 

Note that such dependence of  the numerical value of the gluon mass
condensate on the number of perturbative terms accounted in the corresponding
OPE gives an extra support to the the hypothesis of duality between
perturbative and non-perturbative contributions (see a very recent
work \cite{Narison:2009ag} and references therein).

Finally, we hope that our results will be of use for better understanding of
the present and future data coming from lattice simulations of QCD
propagators.

\vspace{1cm}

\noindent 
{\bf Acknowledgments.}

\noindent 
The authors are grateful to J.~H.~K\"uhn for discussions.  We thank
J. Micheli, A. Le Yaouanc,  O. P\`ene and V.I.  Sakharov for reading
the manuscript and useful advice.

This work is supported by DFG through SFB/TR~9. A.M thanks the
Landesgraduiertenf\"orderung for support.

\appendix

\section{Several massive quarks}
\label{sec:mass}

Until now, all OPEs in this work have been formulated for the case of
(at most) one massive quark with mass $m$. From a physical point of view, this is
a valid approximation: corrections from $u$ and $d$ masses are
in general negligible, while all other quark masses show a strong
hierarchy.

Nevertheless, it is also possible to generalize our results to the case
of $n_f$ massive quarks with masses $m_1,\,m_2,\,\dots,m_{n_f}$. The generalization of
the OPEs of the gluon and ghost propagators is straightforward: We
replace $m^2$ by the sum over $m_i^2$:
\begin{equation}
  \label{eq:OPE_massive_g_h}
  D^?(Q^2) = D_0^?(\mu/Q,a_s)+\frac{C^?_{A^2}(\mu/Q,a_s)}{Q^2} \langle
  A^2 \rangle + \sum_{i=1}^{n_f} \frac{C^?_{m_i^2}(\mu/Q,a_s)}{Q^2}  m_i^2\,,
\end{equation}
where $C^?_{m_i^2}$ is the same as $C^?_{m^2}$ from Eqs. \eqref{eq:c_g_m2} and \eqref{eq:c_h_m2}.

The case of the quark propagator is a bit more complicated.
Without loss of
generality we assume that the external quark has the mass $m_1$. The
correspondingly generalized OPE for the dressing functions $V(Q)$ and $S(Q)$ defined in
Eq. \eqref{quark_prop} then read
\begin{equation}
  \label{eq:OPE_V_massive}
  V(Q)=V_0(\mu/Q,a_s)+
  \sum_{i=1}^{n_f}
  \frac{C^q_{m_i^2}(\mu/Q,a_s)}{Q^2}m_i^2+\frac{C^q_{A^2}(\mu/Q,a_s)}{Q^2}\langle A^2\rangle
\end{equation}
and 
\begin{equation}
  \begin{split}
    \label{eq:OPE_S_massive}
    S(Q)=&\,S_0(\mu/Q,a_s) m_1+ \sum_{i=1}^{n_f}
    \frac{C^q_{m_i^2m_1}(\mu/Q,a_s)}{Q^2}m_i^2m_1\\
    &+\frac{C^q_{m_1A^2}(\mu/Q,a_s)}{Q^2}\langle
    m_1A^2\rangle+\frac{C^q_{\bar{\psi}\psi}(\mu/Q,a_s)}{Q^2}\langle\bar{\psi}\psi\rangle\,.
  \end{split}
\end{equation}
Note that for $i \neq 1$ only even powers of $m_i$ may appear in the
OPEs.

The Wilson Coefficients $C^q_{m_1^2}$, $C^q_{m_1^3}$ and $C^q_{m_1A^2}$ are
identical to $C^q_{m^2}$, $C^q_{m^3}$ and $C^q_{mA^2}$ as given in
Eqs. \eqref{eq:c_q_m2}, \eqref{eq:c_q_m3} and \eqref{eq:c_q_mA2}. 
For $i \neq 1$, the anomalous dimension matrix of \eqref{eq:gam_dim3}
has to be extended to account for the mixing of $\bar{\psi}\psi$ with
$m_i^2m_1$. The corresponding anomalous dimension
$\gamma_{\bar{\psi}\psi,m_i^2m_1}$ can be found in Ref. \cite{Chetyrkin:1994ex}.
The results for the new coefficient functions are (note that below $i \not= 1$!)
\begin{align}
  \label{eq:C_mi2}
  C^q_{m_i^2}= &\,a_s^2\left(\frac{5}{6}-\frac{1}{2}L_Q\right)\notag\\
 &+a_s^3\left(\frac{63659}{3456}-\frac{2243}{192}L_Q+\frac{129}{64}L_Q^2-\frac{5}{2}\zeta_5+\frac{3}{8}\zeta_4\right.\notag\\
 &\quad\left.-\frac{1}{8}\zeta_3-\frac{3}{4}\zeta_3L_Q-\frac{37}{72}n_f+\frac{13}{36}n_fL_Q-\frac{1}{12}n_fL_Q^2\right)\,,\displaybreak[0]\\
  C^q_{m_i^2m_1}=&\,a_s^2\left(\frac{3}{2}\right)+a_s^3\left(\frac{17143}{576}-\frac{319}{32}L_Q\notag\right.\\
 &\quad\left.-5\zeta_5+\frac{41}{4}\zeta_3-\frac{3}{4}n_f+\frac{1}{2}n_fL_Q\right)\,.\displaybreak[0]
\end{align}

\ice{


}

\end{document}